\documentclass[prx,twocolumn,floatfix,superscriptaddress]{revtex4-2}
\usepackage{dcolumn,amsmath}
\usepackage{graphicx}
\usepackage{braket}
\usepackage{bm}
\usepackage{amssymb}
\usepackage{hyperref}
\usepackage{multirow}
\usepackage{footnote}
\usepackage{ragged2e}
\usepackage{xcolor}
\usepackage{pdfcomment}
\usepackage{mathrsfs}
\usepackage{threeparttable,booktabs} 
\usepackage{diagbox}
\usepackage{physics}
\usepackage{tabularx}
\usepackage{multirow}

\DeclareUnicodeCharacter{2212}{-}

\newcommand{\SSD}{Sensor Science Division, National Institute of Standards and Technology, Gaithersburg, Maryland 20899, USA}
\newcommand{\NISTB}{RF Technology Division, National Institute of Standards and Technology (NIST), Boulder, CO 80305}
\newcommand{\JQI}{Joint Quantum Institute, University of Maryland, College Park, Maryland 20742, USA}
\begin{document}

\title{Blackbody Radiation Noise Broadening of Quantum Systems}

\author{Eric B. Norrgard}
\affiliation{\SSD}
\affiliation{\JQI}
\author{Stephen P. Eckel}
\affiliation{\SSD}
\author{Christopher L. Holloway}
\affiliation{\NISTB}
\author{Eric L. Shirley}
\affiliation{\SSD}

\date{\today}

\begin{abstract}
Precision measurements of quantum systems often seek to probe or must account for the interaction with blackbody radiation.  Over the past several decades, much attention has been given to AC Stark shifts and stimulated state transfer.  For a blackbody in thermodynamic equilibrium, these two effects  are determined by the expectation value of photon number in each mode of the Planck spectrum.  Here, we explore how the photon number variance of an equilibrium blackbody generally leads to a parametric broadening of the energy levels of quantum systems that is inversely proportional to the square-root of the blackbody volume.   We consider the the effect in two cases which are potentially highly sensitive to this broadening: Rydberg atoms and atomic clocks.  We find that even in blackbody volumes as small as 1\,cm$^3$, this effect is unlikely to contribute meaningfully to transition linewidths.
\end{abstract}

\maketitle

Precision spectroscopy of atomic and molecular systems is the basis of numerous metrological applications \cite{McGrew2018, Norrgard2021} and tests of fundamental physics \cite{Parthey2011,Safronova2018} and symmetries \cite{Vanhaecke2007a,Ahmadi2018}.  A correct determination of transition energies requires careful accounting of Stark shifts due to ambient blackbody radiation (BBR).   High accuracy BBR shift calculations account for higher order electric and magnetic multipolar moments \cite{Porsev2006} in the scalar and tensor light shifts \cite{Mitroy2003}.  For optical lattice clocks, shifts induced by the trapping laser must also be considered to attain high accuracy (including shifts which are quadratic and higher order in electromagnetic field energy density, known as hyperpolarizability) \cite{Brown2017}.

In addition to shifting energy levels, 
blackbody radiation can  drive transitions to other levels.  This effect can be considered a non-parametric line broadening, as the BBR-stimulated transition rate $\Gamma^{\rm{BBR}}$ adds to the spontaneous decay rate $\Gamma^{\rm{sp}}$ \cite{Farley1981}.  BBR-stimulated decay is prominent in Rydberg atomic systems (typically $\Gamma^{\rm{BBR}} \gtrsim 10^4$ s$^{-1}$ at room temperature), and has also been observed in trapped molecules \cite{Hoekstra2007,Williams2018}, and molecular ions \cite{chou2019}.  
Recently, we surveyed the BBR-induced state transfer and levels shifts in molecular and atomic Rydberg systems, showing these are highly promising platforms for blackbody thermometry applications \cite{Norrgard2021}.

Here, we consider an additional BBR-induced line broadening mechanism that appears to have been overlooked until now.  For a blackbody at temperature $T$ encompassing a volume $V$, fluctuations in the BBR energy produce a root-mean-square (RMS) deviation $\sigma_{\Delta E_i}$ in the BBR shift $\Delta E_i$.  The broadening is proportional to $\sqrt{T^5/V}$ when the energy differences $\hbar\omega_{ij}$ associated with strongest transitions are all large compared to the thermal energy scale (i.e. $\hbar \omega_{ij} \gg k_{\rm{B}}T$).  Unlike BBR-stimulated transitions, the BBR  noise induces a parametric broadening (i.e. the quantum state remains unchanged).  This effect is general to all quantum systems which are radiatively coupled to a thermal bath.
Here, we derive the effect of BBR noise on a quantum system, and then consider  the size of the effect on two experimental systems, atomic clocks and circular Rydberg atoms. For both systems, we find the BBR noise is too small to  significantly contribute to line broadening under any currently feasible scenario.  

Before deriving the BBR noise broadening, we summarize the relevant photon statistics and AC Stark interaction (see Appendix \ref{app: a} for more details on photon statistics).  
For photons with angular frequency $\omega$, we will use $x~=~\hbar\omega/k_B T$ to write the partition function $Z=(1-e^{-x})^{-1}$.  The $m$th moment of the photon number $n$ is
\begin{equation}
\begin{aligned}
    \langle \hat{n}^m \rangle & = \sum_{n=0}^\infty n^m e^{-nx}
    =\frac{1}{Z}\frac{\partial^m}{\partial(-x)^m} Z(x).
\end{aligned}\end{equation}
The variance in photon number for each BBR mode is $\sigma_{\hat{n}}^2 = \langle \hat{n}^2\rangle -\langle \hat{n} \rangle^2 = e^x/(e^x-1)^2$.  In this work, a hat over a symbol indicates an operator.   

We consider an ideal blackbody of effective volume $V$ and temperature $T$ which surrounds a quantum system in state $i$. In this work, we define the effective length $\ell$ and volume $V$ of the cavity through the mode spacing.  Physically, a blackbody cavity is created through a combination of a resistive wall material,  light trapping geometry \cite{Daywitt1994}, and surface micro-structure \cite{Norrgard2016b, Cantat-Moletrecht2020}. A plane wave incident on a surface with non-zero resistance incurs a phase shift according to the Fresnel equations.  Effectively, these phase shifts increase the cavity size and decrease the mode spacing compared to that of an ideal conductor.  Similarly, surface micro-structure can increase the physical length of each incident mode.  For this work, we will not concern ourselves with these complications and instead compare cavities with equal effective size -- that is, cavities with identical mode spacing -- rather than cavities with equal physical dimensions.  

The AC Stark interaction is characterized by a real and an imaginary component \cite{Mitroy2010,Ovsiannikov2012,Solovyev2015}:
\begin{equation}\label{eq:complex shift}
    \mathscr{E}_i^{\rm{BBR}}= \Delta E_i - \frac{i \hbar}{2} \Gamma_i^{\rm{BBR}}=-\int_0^\infty d\omega \frac{\langle\hat{\mathcal{E}}^2\rangle}{2}\alpha_i^s(\omega) ,
\end{equation}
where $\hat{\mathcal{E}}$ is the electric field operator, $\alpha_i^s(\omega)$ is the scalar polarizability of level $i$
\begin{equation}
    \alpha_i^s(\omega) =\sum_j \frac{\vert \mu_{ij}^z\vert^2}{\hbar} \Big(\frac{1}{\omega_{ij}-\omega-\frac{i}{2}\Gamma_{ij}}+\frac{1}{\omega_{ij}+\omega+\frac{i}{2}\Gamma_{ij}}\Big).
\end{equation}
Here, $\mu_{ij}^{z}$ and $\Gamma_{ij}$ are the $z$ component of the dipole matrix element and the partial decay rate, respectively, between state $i$ to state $j$.  The imaginary part $\Gamma_i^{\rm{BBR}}$ of Eq.\,(\ref{eq:complex shift}) is associated with stimulated state transfer from level $i$ to other levels $j$ of the system:
\begin{equation}\label{eq:stimulated rate}
\Gamma_i^{\rm{BBR}}= \sum_j \frac{{\mu_{ij}^z}^2 \omega_{ij}^3}{ \epsilon_0 \hbar \pi c^3} \langle \hat{n}(\omega_{ij}) \rangle.
\end{equation}
The real part of Eq.\,(\ref{eq:complex shift}) is a shift in the energy of level $i$, given by \cite{Gallagher1979, Farley1981}:
\begin{equation}
\begin{aligned} \label{eq: AC Stark shift}
\Delta E_i 
  &=-P \int_{0}^\infty d\omega  \frac{\hbar \omega^3\langle\hat{n}\rangle}{2\varepsilon_0\pi^2 c^3}\alpha_i^s(\omega)
\end{aligned}
\end{equation}
where 
$P$ denotes the Cauchy principal value.  

In the absence of other broadening mechanisms, the interaction of a quantum system with the mean  BBR field leads to a Lorentzian lineshape with full width at half maximum (FWHM)
\begin{equation}
    \Gamma_i = \Gamma_i^{\rm{sp}} + \Gamma_i^{\rm{BBR}},
\end{equation}
where $\Gamma_i^{\rm{sp}}$ and $\Gamma_i^{\rm{BBR}}$ are the rates of spontaneous decay and BBR stimulated depopulations for level $i$, respectively.
Here, we seek to quantify the small, additional broadening of the quantum level induced by fluctuations in the BBR electric field.  These fluctuations lead to RMS deviations $\sigma_{\Delta E_i}$ in the AC Stark shift $\Delta E_i$.  These fluctuations occur with typical timescale of the blackbody coherence time $t_c= h/4k_{\rm{B}}T$  ($t_c\,\approx\,40$\,fs at room temperature) \cite{Donges1998}.  Therefore, the expected lineshape for a single measurement with duration $t_m\gg t_c$ is a Voigt profile with Lorentzian width $\gamma=\Gamma_i$ and  Gaussian half width at half maximum (HWHM) $\sigma=\sigma_{\Delta E_i}\sqrt{t_c/t_m}$.  The FWHM of a Voigt profile is approximately
\begin{equation}
    \rm{FWHM_{V}} \approx \frac{\gamma}{2}+\sqrt{\frac{\gamma^2}{4}+8\sigma^2\ln{2}}.
\end{equation}
In the limit $\gamma \gg \sigma$, the Voigt FWHM becomes
\begin{equation}\label{eq: voigt hwhm}
    \rm{FWHM_{V}} \approx\gamma +8\ln{2} \frac{\sigma^2}{\gamma}.
\end{equation}

In order to calculate  $\sigma_{\Delta E_i}$, we begin by considering each photon mode as contributing $\Delta \epsilon_i$ to the total shift, i.e.\ \mbox{$\Delta E_i =\sum_{\rm{modes}} \Delta \epsilon_i$}.  Assuming uncorrelated modes, $\sigma_{\Delta E_i}^2 = \sum_{\rm{modes}} \sigma_{\Delta \epsilon_i}^2$,  it suffices to find the contribution of each mode independently.  The summation  $\sum_{\text{modes}}$ corresponds to  $P\int V D_{\rm{mode}}$ in the continuous limit,
where  $D_{\rm{mode}}$ is the density of modes per unit volume per unit angular frequency $d\omega$,  and \mbox{$D_{\rm{mode}}=D_{\rm{mode}}^{\rm{FS}}=\omega^2 d\omega/\pi^2 c^3$} for free space. 
Combining this with Eq.\,(\ref{eq: AC Stark shift}) yields
\begin{equation}\label{eq: AC Stark shift mode}
\Delta \epsilon_i = -\frac{\hbar \omega \langle \hat{n} \rangle}{2\varepsilon_0 V }\alpha_i^s(\omega), 
\end{equation}
and the variance of the shift is given by
\begin{equation}
\begin{aligned}
\sigma_{\Delta \epsilon_i}^2 &= \left| \frac{\partial\Delta \epsilon_i}{\partial \hat{n}}\right|^2\sigma_{\hat{n}}^2 = \left|  \frac{\hbar\omega}{2\varepsilon_0 V}\alpha_i^s(\omega) \right|^2 \sigma_{\hat{n}}^2.
\end{aligned}
\end{equation}

Finally we add the contributions of each mode in quadrature to arrive at the variance of the AC Stark shift due to BBR noise:

\begin{equation}\label{eq: shift var}
\begin{aligned}
\sigma_{\Delta E_i}^2 &= \sum_{\text{modes}} \sigma_{\Delta \epsilon_i}^2\\
&= P \int_0^\infty  V D_{\rm{mode}} \frac{\hbar^2\omega^2\sigma_{\hat{n}}^2}{4 \varepsilon_0^2 V^2} \Bigg\vert\alpha_i^s(\omega)\Bigg\vert^2\\
&= P \int_0^\infty d\omega \frac{\hbar^2\omega^4}{4 \varepsilon_0^2 \pi^2  c^3 V}\frac{e^{\hbar\omega/k_{\rm{B}}T}}{(e^{\hbar\omega/k_{\rm{B}}T}-1)^2}\\
&\quad\times\Bigg\vert\sum_{j}\frac{1}{\hbar}
\Big(\frac{\vert \mu_{ij}^z\vert^2 }{\omega_{ij}-\omega-\frac{i}{2}\Gamma_{ij}}+\frac{\vert \mu_{ij}^z\vert^2 }{\omega_{ij}+\omega+\frac{i}{2}\Gamma_{ij}}\Big) \Bigg\vert ^2.
\end{aligned}
\end{equation}
where in the last line we take $D_{\rm{mode}}=D_{\rm{mode}}^{\rm{FS}}$. Note that the summation within the modulus in Eq. (\ref{eq: shift var}) is identical to that found in the Kramers-Heisenberg formula for differential scattering cross-section of light (e.g.\ Ref. \cite{Loudon2000} Section 8.7);  that is, we can consider the BBR noise broadening of each level $i$ to be due to variance in BBR Rayleigh scattering rate due to fluctuating photon number in each mode.  Also of note is the RMS fluctuation in the BBR shift $\sigma_{\Delta E_i}$  is proportional to $1/\sqrt{V}$.  While the RMS electric field  of a blackbody  is independent of volume, smaller volumes contain fewer modes which contribute to the field, and thus are subject to larger field variations.  This relationship between BBR fluctuations and volume was first noted by Einstein \cite{Einstein1904}, and Eq.\,(\ref{eq: shift var}) can alternately be derived using standard thermodynamic relations (see Appendix \ref{sec: app c}).

\begin{table*}[t]\caption{BBR Stark shifts and noise broadening for optical clock systems.  All temperature-dependent quantities are evaluated at $T\,=\,300$\,K.  Static differential polarizabilities  ${\alpha_{ij}^s}^\prime(0)$ from Ref.\,\cite{Mitroy2010} are the recommended calculated value, or mean of values when multiple calculations using the highest accuracy method were presented. For measurement time $t_m$, the BBR noise broadening is $\sigma = \sigma_{\Delta E_{ij}^\prime}\sqrt{t_c/t_m}$. }
\begin{tabular}{rlllllllllll}
\hline\hline
&&&&&&&\multicolumn{2}{c}{($V\,=\,10^{-3}$\,m$^{3}$)} &&\multicolumn{2}{c}{($V\,=\,10^{-6}$\,m$^{3}$)}\\
\cline{8-9}\cline{11-12}
&&$\nu_{ij}$\,(PHz)& $y_{ij}$ & $\Gamma^{\rm{sp}}$\,(Hz)  & ${\alpha_{ij}^s}^\prime(0)$ ($a_0^3$)  & $\Delta E_{ij}^\prime$\,(Hz) & $\sigma_{\Delta E_{ij}^\prime}$\,(Hz) &$\sigma_{\Delta E_{ij}^\prime}/\Gamma^{\rm{sp}}$& & $\sigma_{\Delta E_{ij}^\prime}$\,(Hz) &$\sigma_{\Delta E_{ij}^\prime}/\Gamma^{\rm{sp}}$\\ \hline
Ca$^+$ \cite{Mitroy2010} &&0.411 & 65.7  & 1.4$\times 10^{-1}$ & -44.1   & 3.8$\times 10^{-1}$  & 2.75$\times 10^{-5}$ & 1.97$\times 10^{-4}$ && 8.71$\times 10^{-4}$ & 6.22$\times 10^{-3}$ \\
Sr$^+$ \cite{Mitroy2010}&&0.445 & 71.2  & 4.0$\times 10^{-1}$ & -29.3   & 2.5$\times 10^{-1}$  & 1.22$\times 10^{-5}$ & 3.04$\times 10^{-5}$ && 3.84$\times 10^{-4}$ & 9.61$\times 10^{-4}$ \\
Yb$^{+ a}$ \cite{Mitroy2010}&&0.642 & 102.7 & 1.0$\times 10^{-9}$ & 9.3     & -8.0$\times 10^{-2}$ & 1.22$\times 10^{-6}$ & 1.22$\times 10^{3}$ && 3.87$\times 10^{-5}$ & 3.87$\times 10^{4}$ \\
Yb$^{+ b}$ \cite{Mitroy2010}&&0.688 & 110.1 & 3.1  & 42      & -3.6$\times 10^{-1}$ & 2.50$\times 10^{-5}$ & 8.06$\times 10^{-6}$ && 7.90$\times 10^{-4}$ & 2.55$\times 10^{-4}$ \\
Al$^+$ \cite{Mitroy2010}&&1.121 & 179.3 & 8.0$\times 10^{-3}$ & 0.483   & -4.2$\times 10^{-3}$ & 3.30$\times 10^{-9}$ & 4.13$\times 10^{-7}$ && 1.04$\times 10^{-7}$ & 1.31$\times 10^{-5}$ \\
In$^+$ \cite{Mitroy2010}&&1.27  & 203.2 & 8.0$\times 10^{-1}$ & $<\,$30.7    &$>\,$-2.6$\times 10^{-1}$ &$<\,$1.33$\times 10^{-5}$ &$<\,$1.67$\times 10^{-5}$ && $<\,$4.22$\times 10^{-4}$ & $<\,$5.27$\times 10^{-4}$ \\

Lu$^{+}$ \cite{Paez2016,Porsev2018}&& 0.354 & 56.6 & 5.1$\times 10^{-6}$ & 0.059  & -0.00051 & 4.93$\times10^{-11}$ & 9.59$\times10^{-6}$ && 1.56$\times10^{-9}$ & 0.00030\\

Mg \cite{Mitroy2010}&&0.655 & 104.8 & 1.4$\times 10^{-4}$ & 29.9    & -2.6$\times 10^{-1}$ & 1.27$\times 10^{-5}$ & 9.04$\times 10^{-2}$ && 4.00$\times 10^{-4}$ & 2.86  \\
Ca \cite{Mitroy2010}&&0.454 & 72.6  & 5.0$\times 10^{-4}$ & 133.2   & -1.15  & 2.51$\times 10^{-4}$ & 5.02$\times 10^{-1}$ && 7.94$\times 10^{-3}$ & 1.59$\times 10^{1}$ \\
Sr \cite{Mitroy2010}&&0.429 & 68.6  & 1.4$\times 10^{-3}$ & 261.1   & -2.3  & 9.65$\times 10^{-4}$ & 6.89$\times 10^{-1}$ && 3.05$\times 10^{-2}$ & 2.18$\times 10^{1}$ \\
Yb \cite{Mitroy2010}&&0.518 & 82.9  & 8.0$\times 10^{-3}$ & 155     & -1.33  & 3.40$\times 10^{-4}$ & 4.25$\times 10^{-2}$ && 1.08$\times 10^{-2}$ & 1.34  \\
Zn \cite{Mitroy2010}&&0.969 & 155.0 & 2.5$\times 10^{-3}$ & 29.57   & -2.6$\times 10^{-1}$ & 1.24$\times 10^{-5}$ & 4.95$\times 10^{-3}$ && 3.91$\times 10^{-4}$ & 1.57$\times 10^{-1}$ \\
Cd \cite{Mitroy2010}&&0.903 & 144.5 & 1.0$\times 10^{-2}$ & 30.66   & -2.6$\times 10^{-1}$ & 1.33$\times 10^{-5}$ & 1.33$\times 10^{-3}$ && 4.21$\times 10^{-4}$ & 4.21$\times 10^{-2}$ \\
Hg \cite{Mitroy2010}&&1.129 & 180.6 & 1.1$\times 10^{-1}$ & 21      & -1.81$\times 10^{-1}$ & 6.24$\times 10^{-6}$ & 5.68$\times 10^{-5}$ && 1.97$\times 10^{-4}$ & 1.80$\times 10^{-3}$ \\
Tm \cite{Golovizin2019}&&0.263 & 42.1  & 1.2  & -0.063  & 5.4$\times 10^{-4}$  & 5.62$\times 10^{-11}$ & 4.68$\times 10^{-11}$ && 1.78$\times 10^{-9}$ & 1.48$\times 10^{-9}$ \\
W$^{13+}$ \cite{Nandy2016}&&0.539 & 86.2  & 1.5$\times 10^{1}$ & -0.024  & 2.1$\times 10^{-4}$  & 8.16$\times 10^{-12}$ & 5.54$\times 10^{-13}$ && 2.58$\times 10^{-10}$ & 1.75$\times 10^{-11}$ \\
Ir$^{16+}$ \cite{Nandy2016}&&0.750  & 120.0 & 4.0$\times 10^{1}$ & -0.01   & 8.6$\times 10^{-5}$  & 1.42$\times 10^{-12}$ & 3.59$\times 10^{-14}$ && 4.48$\times 10^{-11}$ & 1.13$\times 10^{-12}$ \\
Pt$^{17+}$ \cite{Nandy2016}&&0.745 & 119.2 & 3.9$\times 10^{-11}$ & 0.182   & -1.57$\times 10^{-3}$ & 4.69$\times 10^{-10}$ & 1.19$\times 10^{1}$ && 1.48$\times 10^{-8}$ & 3.77$\times 10^{2}$ \\
$^{229}$Th \cite{Campbell2012,Seiferle2019}& &2.002     & 319.9 & 6.3$\times 10^{-3}$ & $<\,$4$\times 10^{-5}$ & $>\,$-3.4$\times 10^{-7}$ & $<\,$2.27$\times 10^{-17}$ & $<\,$3.61$\times 10^{-15}$ && $<\,$7.16$\times 10^{-16}$ & $<\,$1.14$\times 10^{-13}$ \\
\hline\hline
\end{tabular}\label{tab:clocks}
\begin{tablenotes}                                                                                  \footnotesize
\centering
\item[\emph{a}]{
$a- 4f^{14}6s-4f^{13}6s^2\ ^2F_{7/2}$.
\quad\quad\quad\quad
$b- 4f^{14}6s-4f^{14}5d\ ^2D_{3/2}$.
}
\end{tablenotes}

\end{table*}

Often, the quantum observable of interest is not the AC Stark shift, but the differential AC Stark shift \mbox{$\Delta E_{ij}^\prime =   \Delta E_j-\Delta E_i$}, such as when measuring the transition frequency $\omega_{ij}$ between two states $i,j$.  Conventionally, negative shifts imply a decrease in the observed transition frequency. Likewise, for the BBR noise line broadening, we define the differential shift per mode \mbox{$\Delta \epsilon_{ij}^\prime = \Delta \epsilon_j - \Delta \epsilon_i$}:
\begin{equation}\label{eq: diff AC Stark shift mode}
\begin{aligned}
\Delta \epsilon_{ij}^\prime &=\frac{\hbar\omega \langle \hat{n} \rangle}{2\varepsilon_0 V }{\alpha_{ij}^{s}}^\prime(\omega).
\end{aligned}
\end{equation}
with ${\alpha_{ij}^{s}}^\prime(\omega)=\alpha_{j}^s(\omega)-\alpha_{i}^s(\omega)$ the differential dynamic scalar polarizability.
 
In many cases, is it appropriate to make a static approximation $y_{ij} = \hbar\omega_{ij}/k_{\rm{B}}T\,\gg\,1$.  In this limit,
\begin{equation}\label{eq:shift static}
    \Delta E_{ij}^\prime = \frac{\hbar}{2\varepsilon_0  \pi^2 c^3}\frac{\pi^4}{15} \Big(\frac{k_{\rm{B}}T}{\hbar}\Big)^4 {\alpha_{ij}^{s}}^\prime(0),
\end{equation}
\begin{equation}\label{eq:var static}
    \sigma_{\Delta E_{ij}^\prime}^2 = \frac{\hbar^2}{4\varepsilon_0^2  \pi^2  c^3 V}\frac{4\pi^4}{15} \Big(\frac{k_{\rm{B}}T}{\hbar}\Big)^5 \Big\vert{\alpha_{ij}^{s}}^\prime(0)\Big\vert^2,
\end{equation}
where we have used the identities
\begin{equation}
    \int_0^\infty dx \frac{x^3}{e^x-1} = \frac{\pi^4}{15},  \quad \int_0^\infty dx \frac{x^4e^x}{(e^x-1)^2} = \frac{4\pi^4}{15}.
\end{equation}

To the best of our knowledge, the BBR noise broadening effect described by Eq.\,(\ref{eq: shift var}) has not been observed experimentally.
That is unsurprising, as in most cases, this effect is quite small. 
If we consider a ``typical'' atomic transition to have frequency \mbox{$\omega_{ij}\,=\,10^{15}$\,s$^{-1}$} and  differential polarizability \mbox{${\alpha_{ij}^{s}}^\prime\,=\,4\pi\varepsilon_0\times1\,a_0^3$} (where $a_0\!\approx\!52.9$\,pm is the Bohr radius), then  $\sigma\!=\!\sigma_{\Delta E_{ij}^\prime}\sqrt{t_c/t_m}\approx\!3.0\times10^{-16}\,\sqrt{t_c T^5/t_m V}$\,Hz\,m$^{3/2}$\,K$^{-5/2}$.  For $T\,=\,300$\,K and $V\,=\,1$\,L\,=\,$10^{-3}$\,m$^3$, $\sigma_{\Delta E_{ij}^\prime}\sqrt{t_c/t_m}\,\approx\,h\times 14$\,nHz.  In nearly all practical measurements, the fluctuation  $\sigma\,=\,\sigma_{\Delta E_{ij}^\prime}\sqrt{t_c/t_m}$   averages down the numerically small value of $\sigma_{\Delta E_{ij}^\prime}$  by a considerable factor (e.g.\ for $t_m = 1$\,s, $\sqrt{t_c/t_m} \approx 2\times 10^{-7}$ at room temperature). We then see using Eq.\,\eqref{eq: voigt hwhm} that the BBR noise broadening becomes a vanishingly small correction to the the linewidth ($\sigma\approx\!3$\,fHz in this example).

We detail below two special cases where BBR noise broadening might be most noticeable: atomic clocks and circular Rydberg atoms.  However, we find in both cases the BBR noise is too small to be detected with modern experimental techniques.

In Table \ref{tab:clocks} we consider several current and proposed atomic frequency standards at $T\,=\,300$\,K.   We calculate the RMS differential BBR shift deviation $\sigma_{\Delta E_{ij}^\prime}$ using Eq.\,(\ref{eq:var static}) as well as the differential BBR Stark shift $\Delta E_{ij}^\prime$ using Eq.\,(\ref{eq:shift static}) as a check of consistency with the atomic transition data references \cite{Mitroy2010,Nandy2016,Campbell2012,Seiferle2019}.  
For the BBR  shift of optical clock transitions, the static approximation is generally accurate to within a few percent \cite{Mitroy2010}.  Differential polarizablities in Table \ref{tab:clocks} are given in their respective references in atomic polarizability units.  These may be converted to SI units (Hz/(V/m)$^2$) by multiplying by a factor of $4\pi\varepsilon_0 a_0^3/h$, with $a_0$ the Bohr radius.

\begin{figure*}
    \centering
     \includegraphics[width=\textwidth]{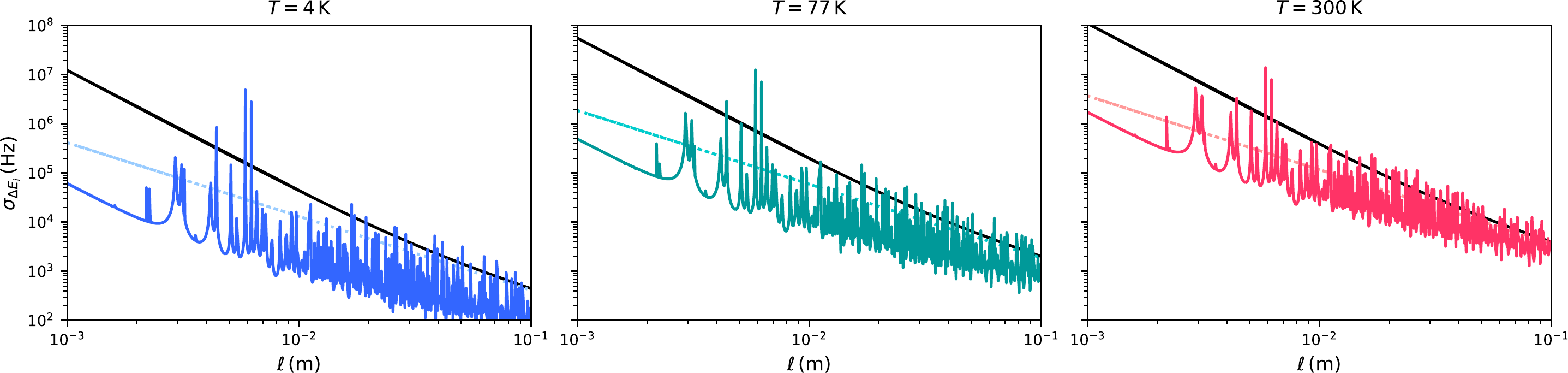}
    \caption{Blackbody radiation noise $\sigma_{\Delta E_i}$ as a function of the length $\ell$ of a cubic cavity.  From left to right, panels show the circular state of Rb with principal quantum number $n=52$ for cavity temperature 4\,K, 77\,K, and 300\,K, respectively.  The dashed, solid black, and solid colored lines calculate $\sigma_{\Delta E_i}$ with the density of modes $D_{\rm{mode}}$ equal to free space, perfectly absorbing walls, and copper walls, respectively. }
    \label{fig: noise vs length}
\end{figure*}

The results of Table I show that detection of BBR noise broadening in atomic frequency standards is not possible without several orders of magnitude improvement over current frequency sensitivity.  Consider as examples the Sr and Yb systems, which now routinely achieve fractional uncertainty of $\delta \omega/\omega \sim 10^{-18}$ \cite{Beloy2014,Nicholson2015,Ushijima2015}. Ignoring numerous technical noise sources (e.g. lattice phonon scattering in optical lattice clocks \cite{Dorscher2018}), even in an exceptionally small BBR volume $V\,=\,10^{-6}$\,m$^3$ at $T\,=\,300$\,K, $\sigma_{\Delta E_{ij}^\prime}$ is only approximately 30\,mHz in Sr and 10\,mHz in Yb.  For a measurement time $t_m\,=\,1$\,s, the BBR noise broadening is then $\sigma \approx 6$\,nHz for Sr and  $\sigma \approx 2$\,nHz for Yb.  Using Eq.\,\eqref{eq: voigt hwhm}, the observed FWHM would then exceed $\Gamma^{\rm{sp}}$ by approximately $(1\times10^{-10})\Gamma^{\rm{sp}}$ for Sr, or $(4\times10^{-13})\Gamma^{\rm{sp}}$ for Yb.
The small differential polarizability characteristic of ions \cite{Mitroy2010,Paez2016,Nandy2016}, inner-shell transitions \cite{Golovizin2019,Nandy2016}, and the nuclear transition of $^{229}$Th \cite{Campbell2012,Seiferle2019} makes the  BBR noise broadening in these systems even smaller than the Sr and Yb cases.

We next consider the possibility of observing BBR noise-limited linewidths in circular Rydberg states $\ket{n\rm{C}}\equiv \ket{n, L=n-1,J=n-1/2}$.  Here we use the Alkali Rydberg Calculator python package \cite{ARC2020} to calculate transition matrix elements and energies for circular states of Rb.  The largest transition dipole matrix elements for Rydberg atoms with principal quantum number $n$ are to states with $n^\prime=n\pm1$.  For our  circular Rydberg state calculations, we include electric dipole transitions to $n^\prime=n-1,n+1,n+2,n+3$.

Rydberg transition wavelengths may be as large or larger than the length scale of its surroundings, and we must consider  cavity effects on the mode density $D_{\rm{mode}}$.    Figure \ref{fig: noise vs length} shows $\sigma_{\Delta E_i}$ for the $\ket{52\rm{C}}$ state of Rb in an effective cubic volume $V\,=\,\ell^3$ for three key conceptual cases.  The dashed colored lines depict  $\sigma_{\Delta E_i}$ calculated using $D_{\rm{mode}}= D_{\rm{mode}}^{\rm{FS}}$; in this case $\sigma_{\Delta E_i}$ is strictly proportional to $\ell^{-3/2}$.  The solid black line depicts an ideal blackbody (i.e.\ perfectly absorbing walls); the mode density for a blackbody $D_{\rm{mode}}^{\rm{BB}}$ is found by quantizing the cavity modes in the usual manner and assigning each mode a finesse $\mathscr{F}\,=\,1/4$ (see Appendix \ref{app: b} for additional details on cavity effects on mode density).  Finally, solid colored lines depict a cubic copper cavity.   The mode density for copper $D_{\rm{mode}}^{\rm{Cu}}$ is complicated by the fact that the resistivity $\rho$ of cryogenic copper may vary by two orders of magnitude depending on purity \cite{Ekin2006}; we assume residual-resistance ratios typical of oxygen-free high conductivity Cu: $\rho(T=300\,\rm{K})/\rho(T=77\,\rm{K}) =10$ and $\rho(T=300\,\rm{K})/\rho(T=4\,K) =100$, with $\rho(T=300\,\rm{K})=1.7\times10^{-8}$\,$\Omega\cdot$m.

\begin{figure}[b]]
    \centering
     \includegraphics[width=\columnwidth]{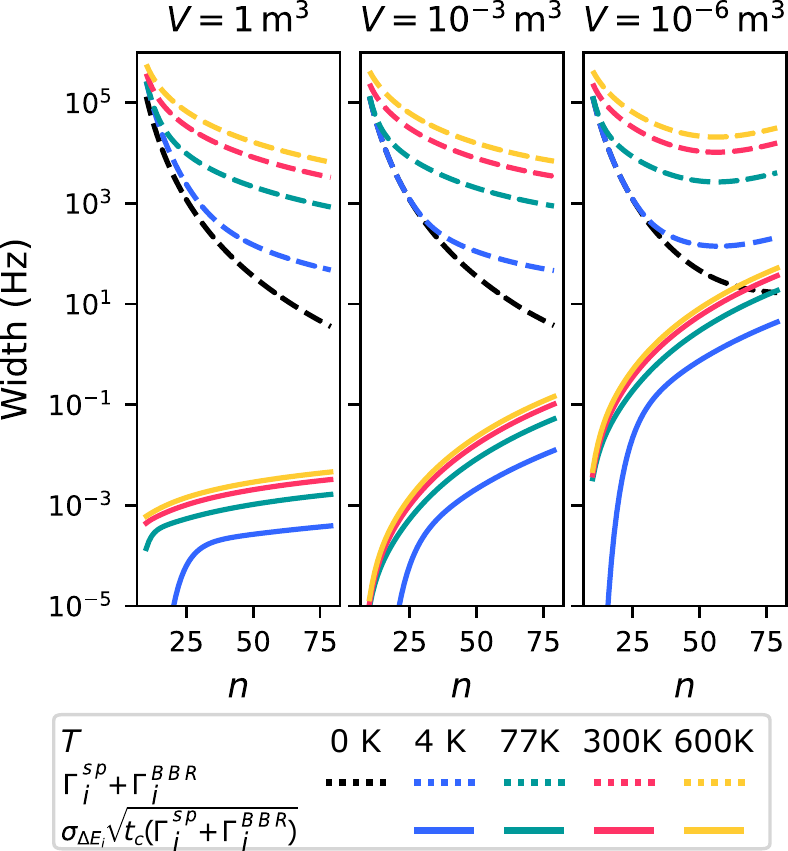}
    \caption{Decay rate $\Gamma^{\rm{sp}}+\Gamma^{\rm{BBR}}$ (dashed lines) and blackbody radiation noise $\sigma_{\Delta E_i}\sqrt{t_c(\Gamma^{\rm{sp}}+\Gamma^{\rm{BBR}})}$ assuming decay rate-limited measurement time (solid lines) for circular states of Rb as a function of principal quantum number $n$. From left to right, panels show a cubic ideal blackbody with effective volume $V$ of  1\,m$^{3}$, $10^{-3}$\,m$^{3}$, and $10^{-6}$\,m$^{3}$, respectively. }
    \label{fig:fig1}
\end{figure}
Figure \ref{fig:fig1} considers Rb circular states for principal quantum number $n \leq 80$ in a cubic ideal blackbody.  The black dashed lines depict the Lorentzian partial linewidth due to spontaneous decay $\Gamma_i^{\rm{sp}}$.  Colored dashed lines depict the partial linewidth due to both spontaneous and BBR-stimulated decay $\Gamma_i\,=\,\Gamma_i^{\rm{sp}} + \Gamma_i^{\rm{BBR}}$.  Solid lines depict BBR noise Gaussian width $\sigma_{\Delta E_i}\sqrt{t_c \Gamma_i}$, where we have assumed a measurement time $t_m\,=\,1/\Gamma_i$.  The magnitude of the BBR noise relative to the decay rate  is increased in cryogenic environments, as $\sigma_{\Delta E_i}$ decreases more slowly than $\Gamma_i^{\rm{sp}}+\Gamma_i^{\rm{BBR}}$ with decreasing temperature.  For effective volumes $V\,=\,1$\,m$^3$,  $\sigma_{\Delta E_i}$ is smaller than $\Gamma_i$ by roughly six orders of magnitude, even at $T\,=\,4$\,K and $n\,=\,80$.  For blackbodies with $V\,=\,10^{-6}$\,m$^3$, $\sigma_{\Delta E_i}$ is smaller than $\Gamma_i$ by roughly two orders of magnitude at $T\,=\,4$\,K and $n\,=\,80$. We caution that for such large $n$, typical relevant transition wavelengths are similar to or exceed $\ell\,=\,1$\,cm for this case.  The assumption of an ideal blackbody for high $n$ and small $V$ is likely invalid, with the BBR noise reduced from these estimates by one or more orders of magnitude as in Fig.\,\ref{fig: noise vs length}. 
The observed FWHM for circular Rydberg states is therefore unlikely to exceed $\Gamma_i$ by more than roughly $10^{-6}\times\Gamma_i$ in the most favorable cases.

In this work, we have assumed isotropic polarization of the BBR, and thus only considered the scalar polarizability.  As each mode has two independent polarizations, polarization fluctuations in the BBR can lead to broadening which involves the vector and tensor polarizabilities as well, and could be considered in future work.

We have derived a parametric broadening, general to all quantum systems, which is due to interactions with fluctuations in a blackbody radiation field. This BBR noise broadening is most significant in applications which involve small effective BBR volumes, large transition dipole moments, and/or high frequency precision.  However, our presented calculations for several atomic clock transitions and for circular Rydberg states of Rb show that BBR noise broadening is typically too small to be detected in these systems with modern sensitivity by at least several orders of magnitude.  These calculations catalogue a novel quantum noise source as being well below current sensitivity for many ongoing experiments, such as atomic frequency standards and quantum sensing experiments using circular Rydberg atoms.  For future experiments with substantially improved frequency precision, this work also sets a benchmark for testing fundamental thermodynamics of photons.

\section*{Acknowledgement}
The authors thank Dazhen Gu, Andrew Ludlow, Kyle Beloy, Michael Moldover, Marianna Safronova, Cl\'{e}ment Sayrin, Wes Tew, and Howard Yoon for insightful conversations, and thank Nikunjkumar Prajapati, Joe Rice, and Wes Tew for careful reading of the manuscript.

\appendix
\section{Photon Statistics}\label{app: a}
The partition function $Z$ for a blackbody radiation mode with frequency $\omega$ is
\begin{equation}
    Z = \sum_{n=0}^\infty e^{-n\hbar\omega/k_B T} =\frac{1}{1-e^{-\hbar\omega/k_B T}}.
\end{equation}
A standard trick to calculate the $m$th moment of the photon number $n$ is
\begin{equation}
\begin{aligned}
    \langle n^m \rangle & = \sum_{n=0}^\infty n^m e^{-nx}\\
    &=\frac{1}{Z}\frac{\partial^m}{\partial(-x)^m} Z(x).
\end{aligned}\end{equation}
The first few moments of the photon number are listed in Table \ref{tab:moments}.

\begin{table}[h]
\caption{First four moments of of the photon number $n$ for blackbody radiation, with $x=\hbar\omega/k_B T$.}
\centering
\begin{tabular}{cc}
\hline\hline
 &$\langle n^m \rangle$\\
    \hline
    $\langle n\rangle$ & $\frac{1}{e^x-1}$\\[2 mm]
    $\langle n^2\rangle$ &$\frac{e^x+1}{(e^x-1)^2}$ \\[2 mm]
    $\langle n^3\rangle$ & $\frac{e^{2x}+4e^x+1}{(e^x-1)^3}$ \\[2 mm]
    $\langle n^4\rangle$ &$\frac{e^{3x}+11e^{2x}+11e^x+1}{(e^x-1)^4}$\\
    \hline\hline
\end{tabular}

\label{tab:moments}
\end{table}

\section{Thermodynamic Derivation} \label{sec: app c}
Here we rederive the main result of the main text, Eq. (\ref{eq: shift var}), from thermodynamic principles.
The total energy $E^{\rm{BBR}}(\omega)$ contained within the volume $V$ of a blackbody cavity per unit angular frequency is
\begin{equation}
E^{\rm{BBR}}(\omega) = V \frac{\hbar \omega^3}{\pi^2 c^3}\frac{1}{e^{\hbar \omega/k_{\rm{B}}T}-1}.
\end{equation}
The variance of $E^{\rm{BBR}}(\omega)$ is given by
\begin{equation}
\begin{aligned}
    \sigma_{E^{\rm{BBR}}(\omega)}^2 &= k_{\rm{B}}T^2\Big(\frac{\partial E^{\rm{BBR}}(\omega) }{\partial T}\Big)\\
&=V \frac{\hbar^2 \omega^4}{\pi^2 c^3} \frac{e^{\hbar \omega/k_{\rm{B}}T}}{(e^{\hbar \omega/k_{\rm{B}}T}-1)^2}\\
&=V \frac{\hbar^2 \omega^4}{\pi^2 c^3} \sigma_{n(\omega)}^2.
\end{aligned}
\end{equation}
Because the spectral energy density $U(\omega)=\varepsilon_0 \mathcal{E}^2 =E^{\rm{BBR}}(\omega)/V$, the variance of the spectral energy density is given by
\begin{equation}
    \sigma_{U(\omega)}^2 =  \frac{\hbar^2 \omega^4}{V \pi^2 c^3} \sigma_{n(\omega)}^2.
\end{equation}
Note that the fluctuations in $U(\omega)$ are inversely proportional to the volume of the cavity.

Since
\begin{equation}
\begin{aligned} 
\Delta E_i =-\int_0^\infty d\omega \frac{\langle U(\omega)\rangle}{2\varepsilon_0}\alpha_i^s(\omega) ,
\end{aligned}
\end{equation}
then
\begin{equation}
\begin{aligned} 
\sigma_{\Delta E_i}^2 
 &=P \int_{0}^\infty d\omega\,\Big\vert \frac{\alpha^s(\omega)}{2\varepsilon_0}\Big\vert^2\sigma_{U(\omega)}^2\\
&=P \int_0^\infty d\omega \frac{\hbar^2 \omega^4}{4 \varepsilon_0^2 V \pi^2 c^3}\sigma_{\hat{n}}^2 \vert \alpha^s(\omega)\vert^2,
\end{aligned}
\end{equation}
which matches Eq.\,(11) of the main text.

\section{Treating a Blackbody as an Optical  Cavity}\label{app: b}

Here we derive the characteristic cavity parameters for a blackbody cavity.  For a cavity mode of length $\ell$, the free spectral range is
\begin{equation}
    \Delta f_{\rm{FSR}}=\frac{c}{2\ell}.
\end{equation}
If a photon is emitted into a mode of length $\ell$, the characteristic length traveled is precisely $\ell$, since blackbodies are by definition perfect absorbers.  Therefore, the spectral width of a blackbody cavity mode (Lorentzian HWHM) is 
\begin{equation}
\Delta f_{\rm{cav}}^{\rm{BB}} = \frac{c}{\ell}.
\end{equation}

The finesse of a cavity is defined as $\mathscr{F}= \Delta f_{\rm{FSR}}/2\Delta f_{\rm{cav}}^{\rm{BB}}$.  For an ideal blackbody, apparently all modes have a finesse of $\mathscr{F}= 1/4$.

The resonant frequencies of a cubic cavity with side $\ell$ are  $f_q=q\Delta f_{\rm{FSR}}$.  Here, $q^2=q_x^2+q_y^2+q_z^2$, and $q_x, q_y, q_z$ are the number of nodes in the $x, y, z$ dimension plus 1, excluding the boundaries. Allowed modes have $q_x, q_y, q_z \ge 0$, with at least one of $q_x, q_y, q_z\ge1$. The quality factor of mode $q$ is then $Q \equiv f_q/2\Delta f_{\rm{cav}}^{\rm{BB}} = q/4$.

\begin{figure}[b!]
    \centering
    \includegraphics[width=\columnwidth]{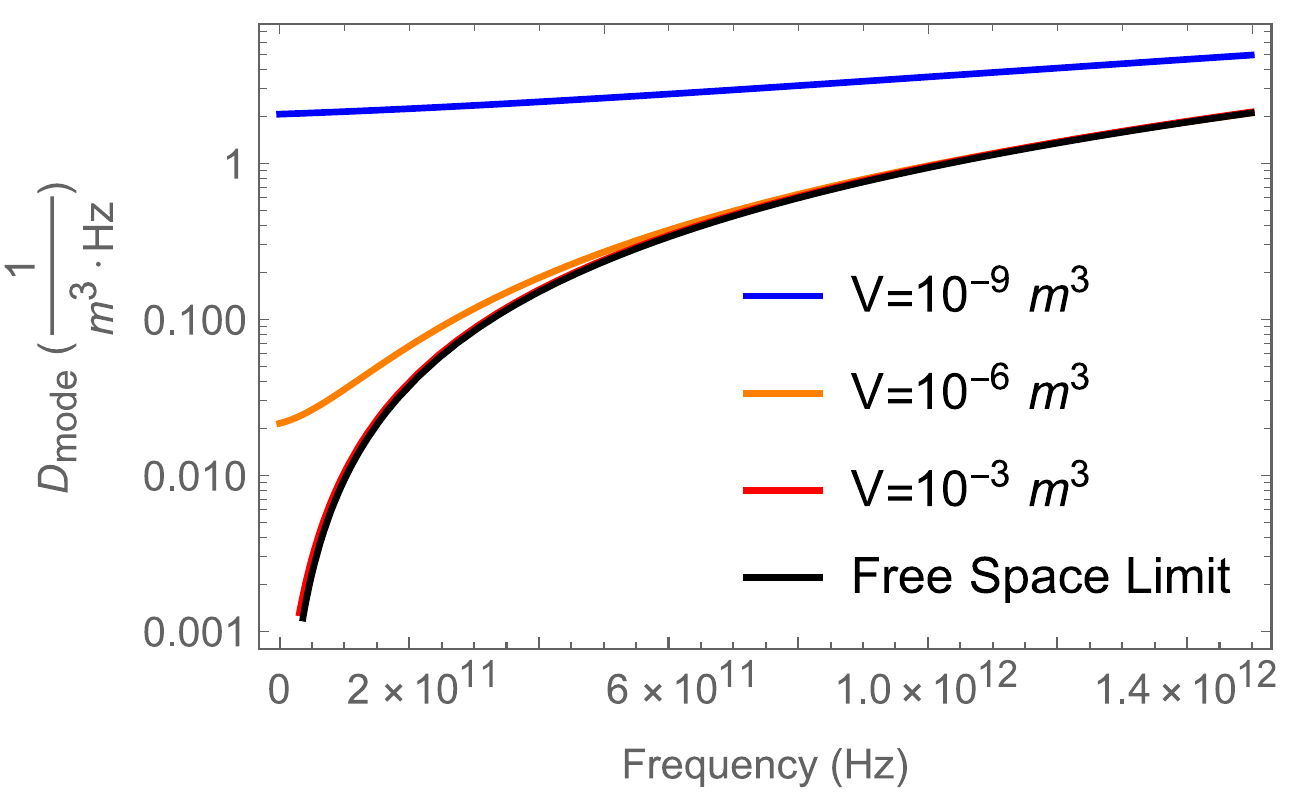}
    \caption{Density of modes $D_{\rm{mode}}$ for a perfect absorber as a function of frequency.  An excess mode density exists at low frequencies for perfect absorbers compared to free space.  }
    \label{fig:my_label}
\end{figure}

Because each mode of the cavity has a spectral width $\Delta f_{\rm{cav}}^{\rm{BB}}$, the spectral density of mode $q$ is
\begin{equation}
d_q(f) = 2\frac{1}{\pi}\frac{f_{\rm{cav}}^{\rm{BB}}}{(f_{\rm{cav}}^{\rm{BB}})^2+(f_q-f)^2} df,
\end{equation}
where the prefactor of 2 accounts for two possible polarizations per mode.  A related quantity is the ``mode density'' 
\begin{equation}
D_{\rm{mode}}(f)= \sum_q d_q(f)/V,
\end{equation}
which is the density of modes per unit frequency per unit volume.  In the limit of $q\gg 1$, the density of modes is approximately a continuous function
\begin{equation}
D_{\rm{mode}}^{\rm{FS}}(f)= \frac{8\pi f^2}{c^3} df.
\end{equation}

For real materials, the finesse will generally have a frequency dependence. One can estimate the quality factor by the relation \cite{Liu1983}
\begin{equation}\label{eq:Q and skin depth}
\begin{aligned}
    Q&= \frac{\rm{Energy\ stored\ in\ cavity}}{\rm{Energy\ lost\ per\ round\ trip}}\\
    &= \frac{3}{2}\frac{V}{S\delta}
\end{aligned}
\end{equation}
where $S$ is the surface area of the cavity ($S=6\ell^2$ for a cube) and $\delta$ is the skin depth given by \cite{Liu1983}
\begin{equation}\label{eq: skin depth}
    \delta=\sqrt{\frac{\varepsilon_0 \rho c^2}{\pi f}}.
\end{equation}
Here $\rho$ is the material's resistivity and $\varepsilon_0$ is the permittivity of free space.  The resistivity $\rho$ is also a frequency dependent pararmeter, but may be approximated by its DC value for sufficiently low frequency.  Under this approximation, we find the finesse of a cubic cavity with walls of resistivity $\rho$ is
\begin{equation}\label{eq: finesse real material}
    \mathscr{F} = \frac{1}{8\sqrt{\varepsilon_0 \rho f/\pi}}.
\end{equation}

\bibliography{thebib}

\end{document}